\documentclass[%
 reprint,
 superscriptaddress,
 amsmath,amssymb,
 aps,
prc,
]{revtex4-2}

\usepackage{graphicx}
\usepackage{dcolumn}
\usepackage{bm}
\usepackage{xcolor} 
\usepackage{hyperref}
\usepackage{float}
\usepackage{placeins}
\usepackage{ulem}
\usepackage{lineno}
\usepackage{booktabs}
\usepackage{soul}
\usepackage{multirow}

\begin{document}

\title{Sequential Clusterization of Light Nuclei and Hypernuclei in Heavy-Ion Collisions within a Wigner Function Coalescence Framework}

\author{Junyi Han}
\affiliation{Key Laboratory of Quark and Lepton Physics (MOE) and Institute of Particle Physics, Central China Normal University, Wuhan 430079, China}
\affiliation{University of Heidelberg, Heidelberg 69120, Germany}

\author{Yue-Hang Leung}
\affiliation{University of Heidelberg, Heidelberg 69120, Germany}

\author{Jiaxing Zhao}
\affiliation{Institute for Theoretical Physics, Johann Wolfgang Goethe University, Frankfurt am Main, Germany}
\affiliation{Helmholtz Research Academy Hessen for FAIR (HFHF),
GSI Helmholtz Center for Heavy Ion Research. Campus Frankfurt, 60438 Frankfurt, Germany}

\author{Yingjie Zhou}
\affiliation{GSI Helmholtzzentrum für Schwerionenforschung GmbH, Planckstr. 1, 64291 Darmstadt, Germany}

\author{Norbert Herrmann}
\affiliation{University of Heidelberg, Heidelberg 69120, Germany}

\author{Yaping Wang}
\thanks{Corresponding author: wangyaping@ccnu.edu.cn}
\affiliation{Key Laboratory of Quark and Lepton Physics (MOE) and Institute of Particle Physics, Central China Normal University, Wuhan 430079, China}

\date{\today}

\begin{abstract}
We investigate the formation of light nuclei and hypernuclei in Au+Au collisions at $\sqrt{s_{NN}} = 3$~GeV within a coalescence framework embedded in the microscopic N-body Parton-Hadron-Quantum-Molecular Dynamics (PHQMD) transport model. The Wigner phase-space distributions employed in the coalescence calculation are constructed from realistic $N$-body wave functions obtained by solving the Schrödinger equation in the hyperspherical harmonics formalism, providing a solid and parameter-free description of nuclear clusters and hypernuclei. By comparing calculated rapidity distributions with STAR data, we extract species-dependent coalescence times, revealing a non-universal formation pattern among different clusters. The resulting yields and kinematic distributions of light nuclei and hypernuclei are systematically analyzed and shown to be sensitive to the underlying wave-function structure and formation time. In addition, we explore cluster–nucleon formation channels for $A=4$ systems. These additional channels improve the description of ${}^{4}\mathrm{He}$ and ${}^{4}_{\Lambda}\mathrm{H}$ yields and help address the underestimation of $A=4$ cluster production in theoretical approaches. Finally, we provide predictions for heavier hypernuclei, including ${}^{5}_{\Lambda}\mathrm{He}$ and ${}^{5}_{\Lambda\Lambda}\mathrm{He}$, which are of interest for future experimental measurements.
\end{abstract}

\maketitle


\section{Introduction}
The production of light nuclei and hypernuclei in heavy-ion collisions involves the formation of loosely bound few-body systems in a rapidly evolving many-body environment. Several mechanisms have been proposed to describe this process, including statistical-thermal models, coalescence at kinetic freeze-out, collision-induced formation via few-body reactions, and potential-driven dynamical formation during the expansion stage~\cite{Oliinychenko2019Snowballs,Oliinychenko2021PionCatalysis,Aichelin1991QMD,Aichelin2020PHQMD,Glassel2022PHQMD,Coci2023DeuteronDynamics,Kireyeu2024PHQMDEoS,Zhang:2025tfd,Andronic2011PLB,Reichert:2023,Kireyeu:2022,Chen:2003,Wang:2026Clustered,Wang:2023Kinetic,Reichert:2022Transport,Staudenmaier:2021,Danielewicz:1991}.
Among these approaches, the coalescence model has been most widely used to describe the formation of nuclei and hypernuclei in heavy-ion collisions.

Experimentally, extensive measurements on the production yields~\cite{STAR2024ProtonLightNuclei3GeV,STARHypernucleiLifetime2022,Chen:2023LightestHypernucleus,Chen:2025Perspectives,ALICEHypernucleiA4,ALICEHypertritonPbPb2025,E864:2002xhb,E864:2000auv,PHENIX:2004vqi,HADES:2025qum,Ono:2019Clusters} and flow~\cite{STAR2022LightNucleiCollectivity,STARHypernucleiFlow2023,ALICE:2024say,HADES:2020lob} of nuclei and hypernuclei have been carried out in relativistic heavy-ion collisions over a wide range of collision energies, providing valuable constraints on the underlying production mechanisms. In particular, collisions at low to intermediate energies ($\sqrt{s_{NN}}\sim$2-10~GeV) provide an unprecedented opportunity to investigate these processes, as cluster production is significantly enhanced in this regime, enabling not only abundant formation of light clusters ($A=2,3$) but also heavier nuclei and hypernuclei. While the coalescence model provides a physically well-motivated description of light-nucleus formation at kinetic freeze-out and successfully reproduces the yields of deuterons, tritons, and $^{3}\mathrm{He}$ over a wide range of collision energies, it significantly underestimates the production yield of $^{4}\mathrm{He}$ in current implementations~\cite{ALICE:2023qyl, Liu:2024ilw}. This discrepancy indicates that important aspects of the cluster formation dynamics may not yet be fully captured, highlighting the need for further developments of the model.

In the Wigner-function coalescence model, the production probability of a composite nucleus in heavy-ion collisions is determined by the phase-space overlap between the Wigner function of the bound-state and the multi-particle phase-space distribution of its constituents~\cite{Scheibl:1998tk}. The invariant yield of an $A$-body nucleus is obtained by integrating the product of the nuclear Wigner density $W_A(\{\mathbf{r}_i,\mathbf{p}_i\})$, constructed from the nuclear wave function, and the multi-particle phase-space distribution of the constituent nucleons and hyperons, integrated over relative coordinates and momenta. The yield is therefore governed by two essential ingredients: the intrinsic structure of the cluster encoded in its wave function, and the phase-space distribution of the coalescing baryons, hereafter referred to as the nucleon source. To extract microscopic information on the production mechanism, one of these ingredients must be constrained as reliably as possible in order to gain insight into the other.

In most practical applications, the Wigner density is approximated by a Gaussian form corresponding to a three-dimensional harmonic oscillator potential~\cite{SatoYazaki1981,ScheiblHeinz1999}. While this approximation is convenient and reasonably justified for compact nuclei, it becomes inadequate for loosely bound systems such as the deuteron and light hypernuclei, which possess extended spatial structures~\cite{Bellini:2020cbj, Mahlein:2025bla, Leung:2025jwe}. In such cases, a Gaussian ansatz introduces an external width parameter that may not reflect the true few-body dynamics, leading to biased coalescence probabilities, particularly for spatially large clusters. Consequently, employing realistic few-body wave functions is not merely a refinement but a necessary step to obtain robust physical insight. Fixing the Wigner density to a realistic form reduces model arbitrariness and allows variations in the calculated yields to be more directly attributed to properties of the nucleon source, such as its size and formation time.

In common coalescence approaches, the ``time'' of coalescence is not treated as a directly observable quantity, but is defined operationally within the modeling framework. In hadronic transport models, nucleons undergo interactions until their last collision, after which they are considered frozen out. The system is evolved up to the cutoff time, and the phase-space coordinates of the nucleons at that time are used to evaluate the coalescence probability. Therefore, in this work, evolution cutoff time serves as an operational definition of the coalescence time $t_{\mathrm{coal}}$.
This prescription is well justified for very loosely bound systems, such as the deuteron, whose small binding energies and large spatial extent render any earlier-formed bound state unstable against hadronic rescattering~\cite{Sun:2022xjr}. However, for more compact and strongly bound nuclei, such as $^{4}\mathrm{He}$, this assumption may be overly restrictive. Owing to their larger binding energies and smaller radii, these nuclei possess higher intrinsic momentum scales and may be less susceptible to disruption by soft late-stage scatterings~\cite{Shuryak:2020yrs}. Moreover, recent studies of baryon number fluctuations indicate that the system exhibits effective multi-baryon correlations that cannot be described by independent-particle dynamics and require attractive higher-order interactions~\cite{Friman:2025swg}. Such correlations suggest that the relevant multi-baryon phase-space structure may already be partially developed prior to the final kinetic freeze-out. As a result, nuclei formation need not be limited to the very last kinetic freeze-out of individual nucleons, motivating a closer examination of the time dependence of coalescence for strongly bound clusters.

In this work, we investigate the formation time of light nuclei and hypernuclei in
Au+Au collisions at $\sqrt{s_{NN}}=3$~GeV within a Wigner-function-based coalescence
framework. The Wigner phase-space densities are constructed from few-body bound-state wave
functions obtained by numerically solving the Schrödinger equation in a
hyperspherical basis. The time dependence of cluster formation is explored by varying the
coalescence time, implemented by applying the coalescence procedure at different
stages of the hadronic evolution in a transport model. The resulting yields and
kinematic distributions are systematically compared with published STAR
measurements at $\sqrt{s_{NN}}=3$~GeV. 

This paper is organized as follows. In Sec.~II, we describe the
construction of the few-body wave functions and the associated Wigner phase-space
densities used in the coalescence calculations, the
transport model setup, and the coalescence procedure, including
the implementation of the time-dependent coalescence criterion. In Sec.~III, we
present a detailed comparison of the model calculations with published
experimental data and discuss the implications for the formation times of light
nuclei and hypernuclei.  Finally, a summary of the main results and an outlook for future studies are given
in Sec.~IV.

\section{Framework}

\subsection{Transport model setup}

Our study is based on the  Parton-Hadron-Quantum-Molecular Dynamics (PHQMD)  microscopic transport approach   \cite{Aichelin:2019tnk,Glassel:2021rod,Kireyeu:2022qmv,Coci:2023daq,Zhou2026EOSClusters}. PHQMD is a microscopic n-body transport model based on the QMD propagation of the baryonic degrees-of-freedom and the dynamical properties and interactions in- and out-of-equilibrium of hadronic and partonic degrees-of-freedom of the Parton-Hadron-String-Dynamics (PHSD) approach \cite{Cassing:2009vt}.
The default clusterization scheme in PHQMD combines the advanced Minimum Spanning Tree (aMST) algorithm and a ``kinetic'' production mechanism. This kinetic mechanism is implemented only for deuterons and incorporates catalytic hadronic reactions, including all isospin channels of the reactions $\pi NN \leftrightarrow \pi d$ and $NNN \leftrightarrow Nd$. 
In the present work, the default clusterization procedure in PHQMD is switched off. Instead, a coalescence model is applied to nucleons and hyperons at a selected time step to study the production of light nuclei and hypernuclei in heavy-ion collisions.

At $\sqrt{s_{NN}} = 3$~GeV, where a high baryon density system is formed, experimental measurements of particle yields and collective flow indicate that baryonic mean-field dynamics play an essential role, and previous studies have shown that their inclusion significantly improves the description of collective flow observables compared to cascade calculations~\cite{Zhou2026EOSClusters,STAR2022DisappearancePartonicCollectivity, STAR2022LightNucleiCollectivity}.
In this work, we therefore employ the mean-field mode of PHQMD with a hard equation of state (EOS), characterized by an incompressibility parameter of $\kappa = 380~\mathrm{MeV}$.

\subsection{Clusterization}
We turn off the default aMST clusterisation in the PHQMD and employ the coalescence mechanism to generate the nuclear clusters and hypernuclei. 
In the present implementation, the transport evolution is terminated at a prescribed evolution time $t_{\mathrm{coal}}$. The phase-space information of each nucleon and hyperon ($\Lambda$) at that time are used as input to the coalescence calculation. 
The probability to form a cluster is determined by the overlap between the phase-space distribution of the constituent baryons in the source and the cluster Wigner function $W_{\mathcal C}$. Specifically, the yield of cluster type $C$ at time $t$ is given by
\begin{eqnarray}
N_{\mathcal C}(t)&=&g_{\mathcal C}\int \prod_{i=1}^N {d^3 {\bf p}_id^3 {\bf r}_i\over (2\pi)^3} F({\bf r}_1,...,{\bf r}_N; {\bf p}_1,...,{\bf p}_N,t) \nonumber\\
&\times &W_{\mathcal C}({\bf r}_i,...,{\bf r}_N; {\bf p}_i,...,{\bf p}_N),
\label{eq.coal}
\end{eqnarray}
where $F$ is the $N$-body phase-space distribution of nucleons or hyperons.
$g_{\mathcal C}=(2J_{\mathcal C}+1)/\prod_i^N (2J_i+1)$ is the spin degeneracy factor, where $J_{\mathcal C}$ and $J_i$ are the total angular momentum of the cluster ${\mathcal C}$ and nucleon (hyperon) $i$, respectively.

The Wigner function of light nuclei and hypernuclei involved in this paper are taken directly from Ref.~\cite{ZhaoAichelinBratkovskaya2025}, obtained by solving the few-body Schr\"odinger equation numerically.
${\bf r}_j,...,{\bf r}_k; {\bf p}_j,...,{\bf p}_k $ are the coordinates of the constituent baryons. For $N$-body system, one usually introduce the Jacobi coordinates to transform the individual particle coordinates to the center-of-mass (CoM) coordinate ${\bf R}$ and the relative coordinate ${\bm \chi}_i$ via,
\begin{eqnarray}
{\bf R}&=& {1\over M}\sum_{i=1}^Nm_i{\bf r}_i, \nonumber\\
{\bm \chi}_{N-j}&=&\sqrt{M_j m_{j+1}\over M_{j+1}\mu }\left( {\bf r}_{j+1}-{1\over M_j}\sum_{i=1}^j m_i {\bf r}_i\right),
\end{eqnarray}
where $M_j=\sum_{i=1}^j m_i$ and $j=1,...N-1$ and $m_i$ is the physical mass of the constituent. $\mu = M=\sum_i m_i$.
The conjugate momentum transformation is:
\begin{eqnarray}
{\bf P}&=&\sum_{i=1}^N {\bf p}_i, \nonumber\\
{\bf q}_{N-j}&=&\sqrt{M_j m_{j+1}\over M_{j+1}\mu }\left( {\mu \over m_{j+1}}{\bf p}_{j+1}-{\mu \over M_j}\sum_{i=1}^j {\bf p}_i\right),
\end{eqnarray}
where ${\bf P}$ and ${\bf q}_i$ are the total and relative momentum, respectively.
The $N$-body motion can be factorized into a CoM motion, which is described by a plane wave, and a relative motion.
Further, the relative coordinates ${\bm \chi}_i$ can be transformed into hyper-radius $\rho\equiv \sqrt{{\bm \chi}_1^2+...+   {\bm \chi}_{N-1}^2}$ and $3N-4$ hyper-angles. The $N$-body Schr\"odinger equation can be solved numerically via the hyperspherical harmonic basis expansion~\cite{Barnea:1999be,Marcucci:2019hml,Zhao:2020nwy}. With this expansion, we found that for the ground state of cluster, the lowest basis dominates, which means the total wave function is almost angles independent. So, we can integrate over all angles to get an angle-independent Wigner function, $W_{\mathcal C}(\rho,q)$. The Wigner function is normalized and the probability distribution can be found in Ref.~\cite{ZhaoAichelinBratkovskaya2025}. The hyper-radius $\rho$ characterizes the spatial extent of the $N$-body system,
while the hyper-momentum $q$ quantifies its internal momentum scale.

In practice, the N-body phase-space distribution $F$ is provided directly from the event-by-event PHQMD output. Consequently, the many-body correlations generated dynamically during the transport evolution, including those arising from the nuclear mean field and hadronic collisions, are naturally preserved. At the selected coalescence time $t$, all baryons within the same event are considered to construct the given N-body configurations. The cluster yield is then evaluated as the event-averaged sum of the coalescence probabilities,
\begin{eqnarray}
N_{\mathcal C}(t)=\left\langle \sum_{i \in p}\sum_{j \in n}...\sum_{N\in \Lambda}g_{\mathcal C}W_{\mathcal C}(\rho^{cm}, q^{cm}) \right\rangle_{\rm event},
\end{eqnarray}
where $\rho^{cm}$ and $q^{cm}$ denote the hyper-radius and hyper-momentum of the constituent baryons, respectively, evaluated in the N-body center-of-mass frame.
In the present implementation, the phase-space distribution is not updated after each coalescence step. Specifically, baryons that contribute to the formation of one cluster remain available for subsequent coalescence calculations. This approximation is adopted for computational simplicity and is justified by the fact that the yields of heavy clusters are much smaller than the total baryon multiplicity in a single event, as demonstrated by the available experimental data at $\sqrt{s_{NN}}=3$~GeV. Although the deuteron yield is comparatively larger, deuterons are predominantly formed at much later stages of the collision because of their extremely small binding energy and large spatial size. Therefore, the depletion of the baryon source due to deuteron formation is also expected to have a negligible effect on the production of heavier clusters.

For ${}^{4}_{\Lambda}\mathrm{H}$, the Wigner function from Ref.~\cite{ZhaoAichelinBratkovskaya2025} corresponds to the ground state. The feed-down contribution from the excited $1^{+}$ state is approximated by assuming the same Wigner function as that of the ground state. The excited state is then included through the spin degeneracy factor~\cite{Bedjidian1976H4Lgamma} and is assumed to decay electromagnetically via ${}^{4}_{\Lambda}\mathrm{H}^{*}(1^{+})\rightarrow {}^{4}_{\Lambda}\mathrm{H}(0^{+})+\gamma$.
We also investigated possible production channels for $A=4$ clusters~\cite{Zhao:2026jek}. The existence and properties of these channels are closely related to the strength and spin dependence of the effective hyperon–nucleon interaction, as well as to the role of three-body forces~\cite{Deltuva:2006sz,Wang:2025ify}. Experimental confirmation of these channels will require future high-statistics measurements. For $^4\mathrm{He}$, relevant two-body channels include $t+p$ and $^3\mathrm{He}+n$. Both satisfy a simple mass criterion, $m_t+m_p \gtrsim m_{^4\mathrm{He}}$ and $m_{^3\mathrm{He}}+m_n \gtrsim m_{^4\mathrm{He}}$. For $^4_\Lambda\mathrm{H}$, there is only one additional channel, namely $^3_\Lambda \mathrm{H}+n$ due to $m_t+m_\Lambda \approx$ $^4_\Lambda\mathrm{H}$ in our previous study~\cite{ZhaoAichelinBratkovskaya2025}. The wave functions and Wigner functions of these cluster–nucleon bound states are obtained by solving the two-body Schrödinger equation with a finely tuned phenomenological two-range Gaussian potential, which reproduces the empirical masses of the $A=4$ hypernuclear system.
For $^5_\Lambda$He and $^5_{\Lambda\Lambda}$He, no experimentally established excited states exist so far, therefore only the ground-state is included.

\subsection{Weighting procedure}
To account for discrepancies between the PHQMD calculations and the experimental proton and $\Lambda$ yields, an empirical weighting procedure is applied. Since light nuclei and hypernuclei are subsequently formed from the nucleons and hyperons generated by the transport calculation, deviations in the proton and $\Lambda$ yields would directly propagate to the predicted yields of composite particles.

For protons, the weight is defined as the ratio of the experimental to model proton yield integrated over the rapidity interval $-0.5<y_{\rm cm}<0$. This mid-rapidity region is chosen minimize contributions from spectator matter.
For the calculation of the proton weight, the experimental inclusive proton yield
\[
p_{\rm incl}^{\rm exp}
= p + d + t + 2\times{}^{3}{\rm He} + 2\times{}^{4}{\rm He}
\]
is used. The proton content carried by hypernuclei is neglected in this correction, since hypernuclear yields are much smaller than those of light nuclei.

For $\Lambda$ hyperons, the weight is determined from the ratio of the experimental to model $\Lambda$ yield integrated over the same mid-rapidity interval, $-0.5<y_{\rm cm}<0$.
The experimental $\Lambda$ yield includes contributions from $\Sigma^{0}\rightarrow\Lambda\gamma$ decays. Accordingly, $\Sigma^{0}$ hyperons are added to the $\Lambda$ yield in PHQMD, and the model $\Lambda$ yield used for the weight calculation contains both directly produced $\Lambda$ hyperons and $\Sigma^{0}$ decay contributions.
In contrast to the proton case, no analogous inclusive correction is applied to the $\Lambda$ yield. This is because the production of hypernuclei, such as ${}^{3}_{\Lambda}\mathrm{H}$ and ${}^{4}_{\Lambda}\mathrm{H}$, constitutes only a small fraction of the total $\Lambda$ yield, typically less than $1\%$.

After clusters are constructed by the Wigner-function coalescence procedure, their yields are weighted by the product of the corresponding constituent weights. The neutron weight is assumed to be equal to the proton weight, $w_n=w_p$, due to the lack of experimental neutron yield measurements. For example, the yield of ${}^{3}_{\Lambda}\mathrm{H}(p+n+\Lambda)$ is weighted by
\[
w_p w_n w_{\Lambda} = w_p^2 w_{\Lambda}.
\]
For the $0-10\%$ centrality class, the proton and $\Lambda$ weighting factors are $w_p=0.998$ and $w_\Lambda=1.029$, respectively. The neutron weight is assumed to be identical to the proton weight, $w_n=w_p$, due to the lack of experimental neutron yield measurements.

With these weighting factors, the triton yield is scaled by $w_p^3=0.994$, corresponding to a reduction of about $0.6\%$, while the hypertriton yield is scaled by $w_p^2w_\Lambda=1.025$, corresponding to an increase of about $2.5\%$.


\section{Results}
\subsection{Last-collision-time distributions}
\begin{figure}[htbp]
    \centering
    \includegraphics[width=0.95\columnwidth]{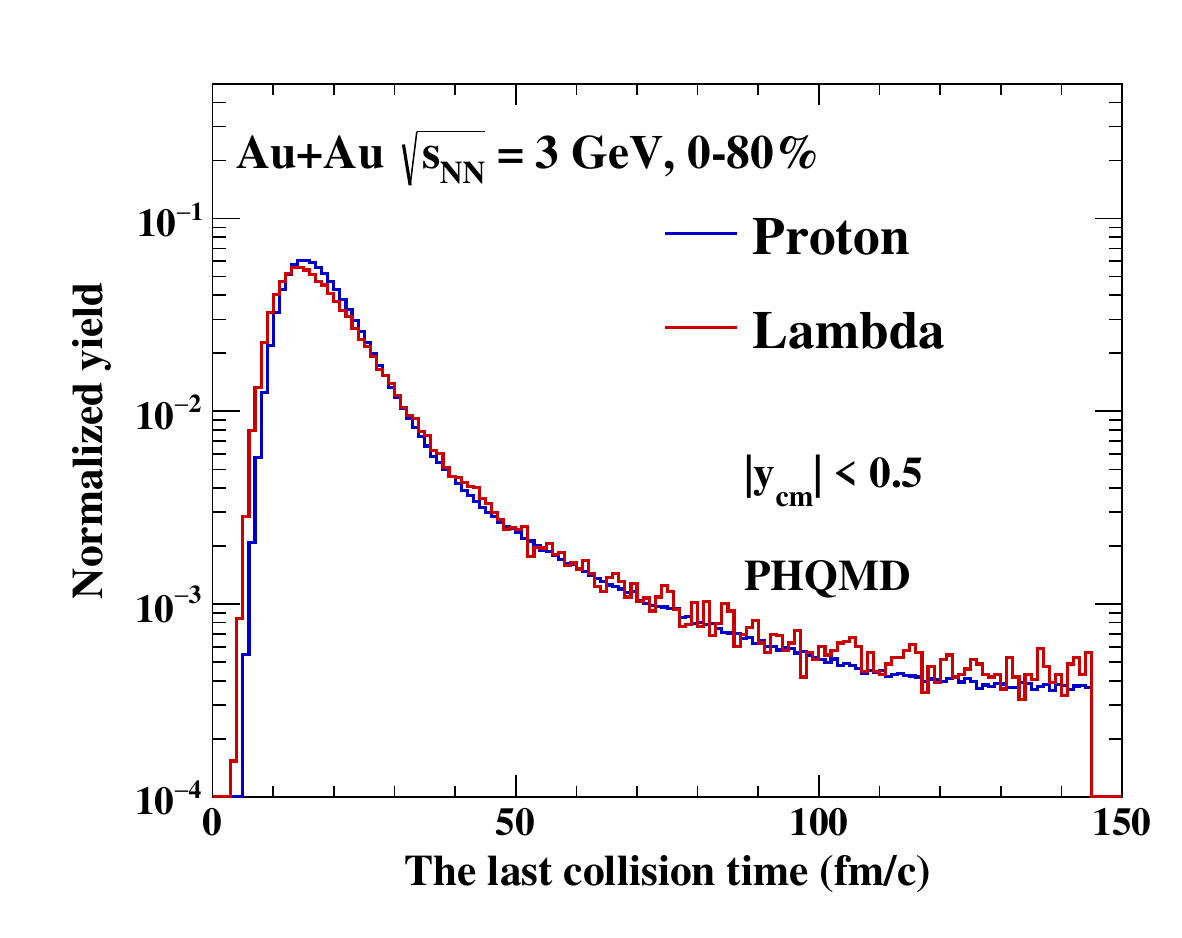}
    \caption{Normalized last-collision-time distributions of protons and $\Lambda$ hyperons in Au+Au collisions at $\sqrt{s_{NN}} = 3$~GeV for the 0--80\% centrality. The distributions are shown for particles within $|y_{\mathrm{cm}}| < 0.5$. All distributions are normalized to unity.}
    \label{fig:fig1_frt}
\end{figure}

In the PHQMD calculations, the centrality is selected according to the impact parameter $b$. Here, the time evolution is followed up to $145~\mathrm{fm}/c$.
Figure~\ref{fig:fig1_frt} shows the normalized last-collision-time distributions of protons and $\Lambda$ hyperons in Au+Au collisions at $\sqrt{s_{NN}} = 3$~GeV for the 0--80\% centrality.
The proton and $\Lambda$ yields shown in the PHQMD calculations are inclusive yields, including baryons that may subsequently participate as constituents in cluster formation.
The last collision time is used to represent the kinetic freeze-out time. 
Both distributions peak at approximately $15~\mathrm{fm}/c$ and then decrease rapidly with increasing time. More than 99\% of protons and $\Lambda$ hyperons freeze out before $145~\mathrm{fm}/c$, indicating that by this time nearly all baryons relevant for cluster formation have undergone their last collision.
Since more than 99\% of protons and $\Lambda$ hyperons freeze out before
$145~\mathrm{fm}/c$, performing coalescence at this time provides an
approximation to the conventional kinetic freeze-out prescription. To explore
the possibility of cluster formation at earlier stages of the hadronic
evolution, the evolution cutoff time $t_{\mathrm{coal}}$ is varied from
20 to 145~fm/$c$.

\subsection{Production of clusters}
We now turn to a systematic comparison between the PHQMD+coalescence calculations and the published STAR measurements at $\sqrt{s_{NN}}=3$~GeV.

\begin{figure}[htbp]
    \centering
    \includegraphics[width=0.95\columnwidth]{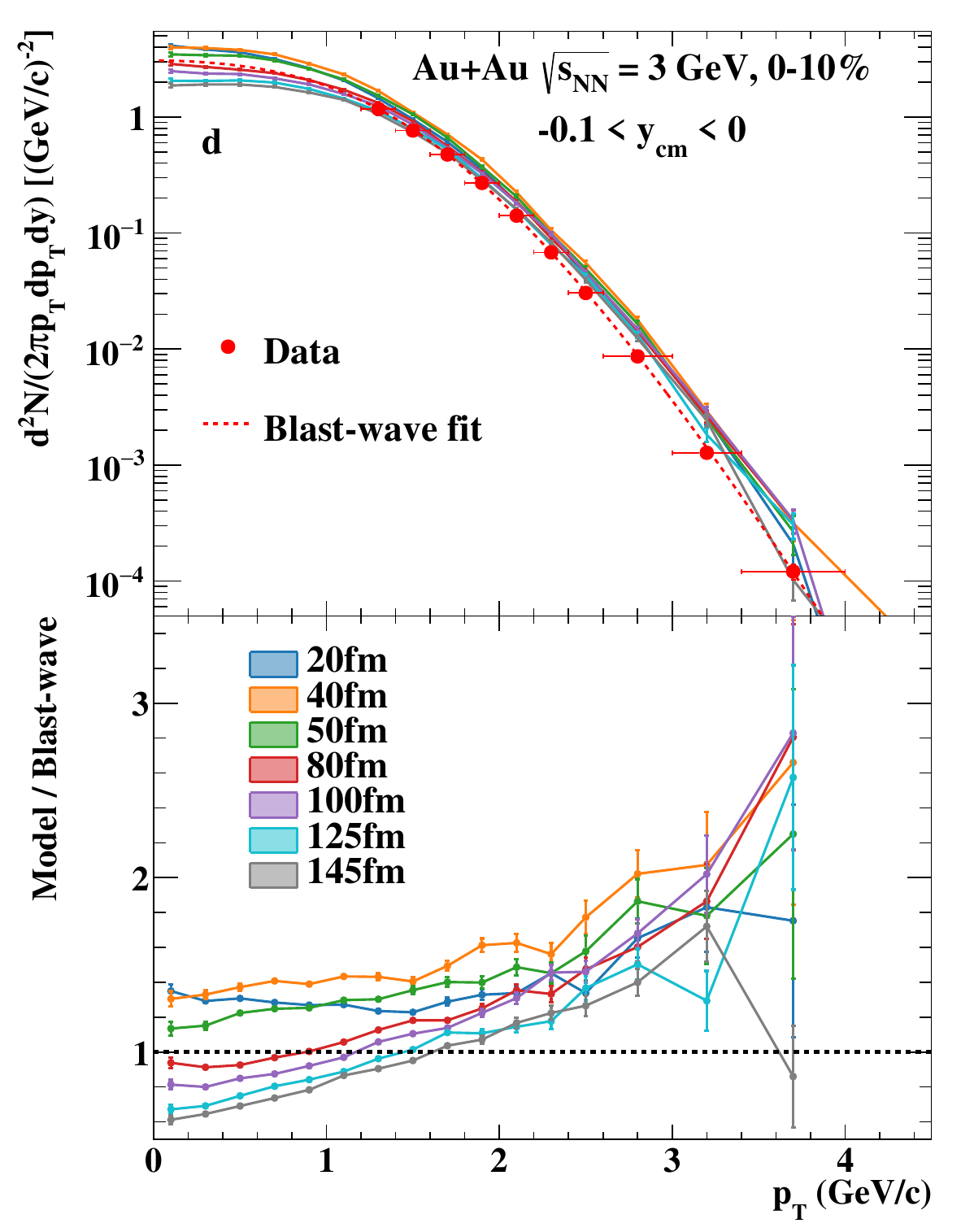}
    \caption{Transverse-momentum spectra of deuterons in Au+Au collisions at $\sqrt{s_{NN}} = 3$~GeV for the 0--10\% centrality and $-0.1<y_{\rm cm}<0$. 
    The red markers represent the STAR experimental data~\cite{STAR2024ProtonLightNuclei3GeV}. The red dashed curve denotes the blast-wave fit to the experimental spectrum, while the colored solid curves show the PHQMD+coalescence results obtained with different $t_{\mathrm{coal}}$. The lower panel shows the ratio of the PHQMD+coalescence results to the blast-wave fit.}
    \label{fig:fig2_pt_d}
\end{figure}

As an example, Figure~\ref{fig:fig2_pt_d} shows the deuteron transverse-momentum spectrum for the 0--10\% centrality in the rapidity interval $-0.1<y_{\mathrm{cm}}<0$. A clear dependence on the evolution cutoff time $t_{\mathrm{coal}}$ is observed. 
Although an earlier $t_{\mathrm{coal}}$ might naively be expected to enhance deuteron production due to the smaller spatial separation between constituents, the yields do not vary monotonically with $t_{\mathrm{coal}}$. In particular, the yield at $t_{\mathrm{coal}}=20~\mathrm{fm}/c$ can be attributed to the incomplete decay of resonances at such early times, which reduces the number of available protons and neutrons entering the coalescence calculation.

As shown in Figure~\ref{fig:fig2_pt_d}, none of the calculations provides a satisfactory description of the deuteron spectrum over the full $p_{\rm T}$ range covered by the experimental measurements. Most of the measured data points are located at relatively high $p_{\rm T}$, whereas the largest differences among the calculations occur in the unmeasured low-$p_{\rm T}$ region. Since the determination of the total yield $dN/dy$ relies on a blast-wave extrapolation into this region, it is instructive to compare the model predictions with the extrapolated spectrum. The blast-wave extrapolation falls within the range spanned by the model calculations at low $p_{\rm T}$, suggesting that the extrapolated contribution to $dN/dy$ is of a reasonable magnitude. In the experimental analysis, the uncertainty associated with the low-$p_{\rm T}$ extrapolation is accounted for through a systematic uncertainty on the $p_{\rm T}$-integrated yield, obtained by varying the fit function used for the extrapolation.
In the following, we therefore focus on the $p_{\rm T}$-integrated yield $dN/dy$. Any preferred value of $t_{\mathrm{coal}}$ extracted from the $dN/dy$ comparison should be regarded as an effective coalescence time that reflects contributions integrated over the full $p_{\rm T}$ range.

\begin{figure}[htbp]
    \centering
    \includegraphics[width=1\columnwidth]{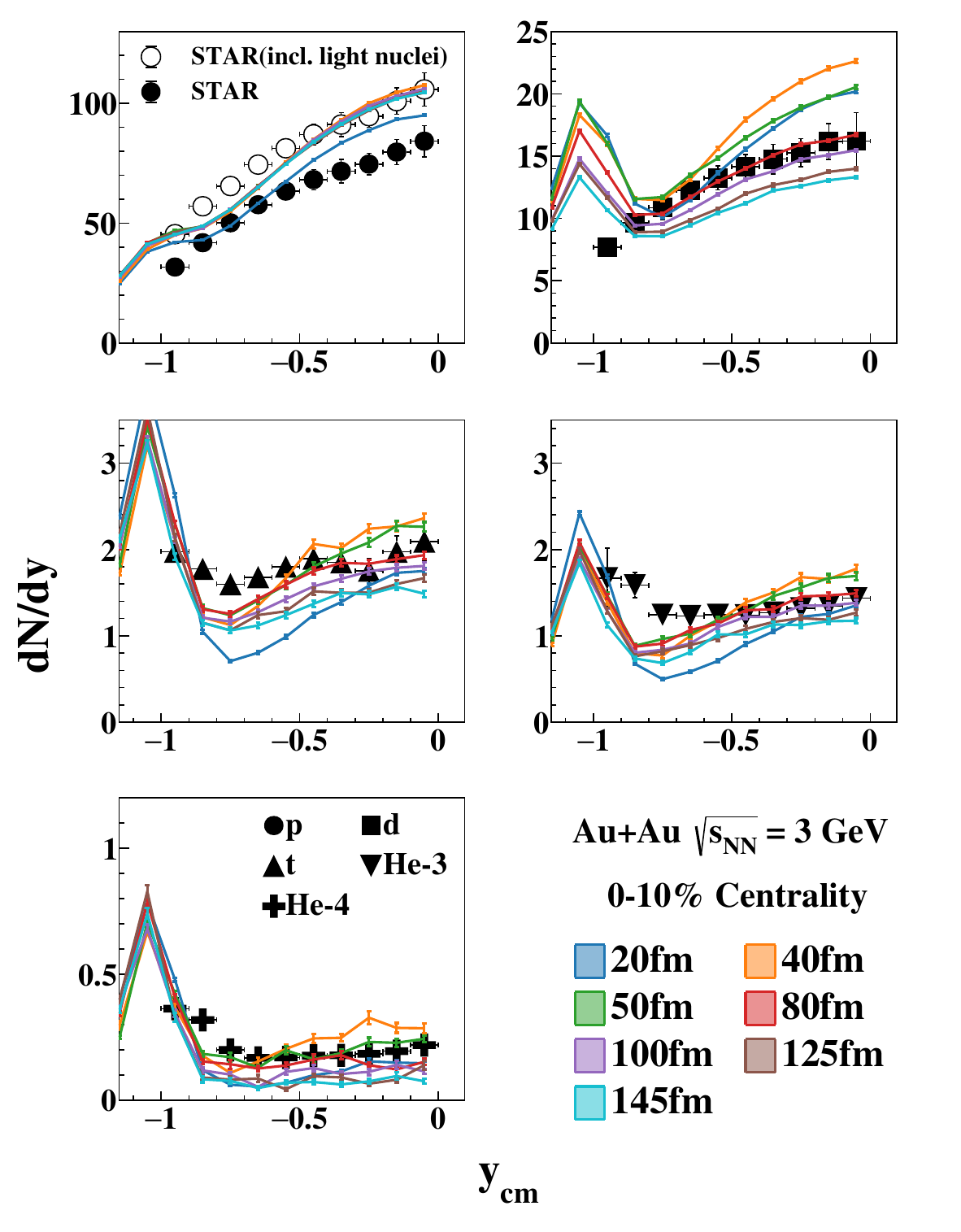}
    \caption{Rapidity distributions, $dN/dy$, of protons, deuterons, tritons, ${}^{3}\mathrm{He}$, and ${}^{4}\mathrm{He}$ in Au+Au collisions at $\sqrt{s_{NN}}=3$ GeV for the 0–-10\% centrality. The curves show the PHQMD+coalescence results obtained with different $t_{\mathrm{coal}}$, while the symbols denote the STAR data. For protons, both the measured proton yield and the inclusive proton yield, including contributions from light nuclei, are shown as indicated in the legend. The ${}^{4}\mathrm{He}$ yield includes contributions from the direct four-body coalescence channel $p+p+n+n$, as well as the sequential channels ${}^{3}\mathrm{He}+n$ and $t+p$.}
    \label{fig:fig4_dndy_LN}
\end{figure}

Figure~\ref{fig:fig4_dndy_LN} illustrates the dependence of the calculated proton, deuteron, triton, ${}^{3}\mathrm{He}$, and ${}^{4}\mathrm{He}$ yields on the coalescence time $t_{\mathrm{coal}}$, together with the corresponding STAR measurements in 0--10\% central Au+Au collisions~\cite{STAR2024ProtonLightNuclei3GeV}.
The PHQMD calculation overpredicts the proton yield around $y_{\rm cm}\sim -1$. Given that the beam rapidity is $y_{\rm beam}\approx 1.05$, this discrepancy occurs near the spectator region. To reduce the influence of spectator-related effects, the following quantitative comparison is restricted to $-0.5<y_{\rm cm}<0$.
Unlike the measured $p_{\rm T}$ spectra, the rapidity distributions involve an integration over the full $p_{\rm T}$ range and are therefore sensitive to the low-$p_{\rm T}$ region not covered by the experimental measurements.
For ${}^{4}\mathrm{He}$, the calculated yield is obtained as the sum of the direct four-body coalescence channel $p+p+n+n$ and the cluster-plus-nucleon channels ${}^{3}\mathrm{He}+n$ and $t+p$.
The dependence of the calculated yields on $t_{\mathrm{coal}}$ reflects the time evolution of the phase-space distributions entering the coalescence calculation.

To quantify the agreement between the model and the experimental data, we calculate the $\chi^2/\mathrm{NDF}$ for each $t_{\mathrm{coal}}$. The definition used here is

\[
\chi^2/\mathrm{NDF}
=
\frac{1}{N_{\mathrm{data}}}
\sum_i
\frac{
\left(
Y_i^{\mathrm{model}}-Y_i^{\mathrm{exp}}
\right)^2
}{
\left(\sigma_i^{\mathrm{model}}\right)^2+
\left(\sigma_i^{\mathrm{exp}}\right)^2
},
\]

\noindent where $Y_i^{\mathrm{model}}$ and $Y_i^{\mathrm{exp}}$ are the model and experimental values in the $i$-th bin of the observable under consideration, respectively. The uncertainties $\sigma_i^{\mathrm{model}}$ and $\sigma_i^{\mathrm{exp}}$ are added in quadrature so that the statistical fluctuation of the model calculation is also taken into account.

\begin{figure}[htbp]
    \centering
    \includegraphics[width=1\columnwidth]{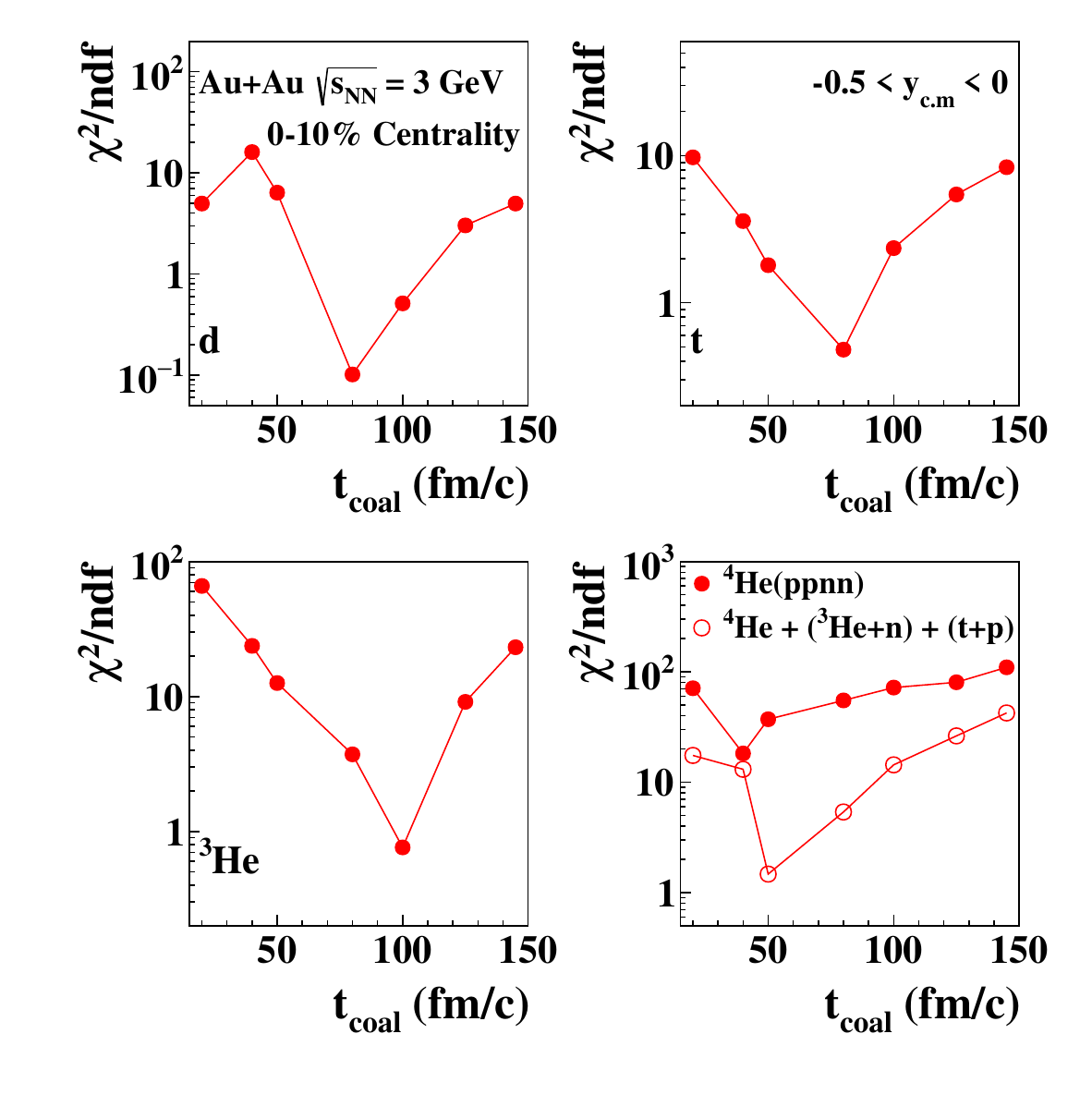}
    \caption{$\chi^2/\mathrm{NDF}$ as a function of the evolution cutoff time $t_{\mathrm{coal}}$ for deuterons, tritons, ${}^{3}\mathrm{He}$, and ${}^{4}\mathrm{He}$ in Au+Au collisions at $\sqrt{s_{NN}}=3$ GeV for the 0--10\% centrality. The $\chi^2/\mathrm{NDF}$ values are calculated by comparing the PHQMD+coalescence rapidity distributions with the STAR data in the rapidity interval $-0.5<y_{\mathrm{cm}}<0$. For ${}^{4}\mathrm{He}$, results are shown for the direct four-body coalescence channel $p+p+n+n$ and for the summed yield including the additional ${}^{3}\mathrm{He}+n$ and $t+p$ channels.}
    \label{fig:fig5_chi2ndf_LN}
\end{figure}

Figure~\ref{fig:fig5_chi2ndf_LN} summarizes the $\chi^2/\mathrm{NDF}$ values obtained from comparisons with the measured rapidity densities in the interval $-0.5<y_{\rm cm}<0$. For deuterons, tritons, and ${}^{3}\mathrm{He}$, the minimum $\chi^2/\mathrm{NDF}$ is found at $t_{\mathrm{coal}}\approx 80$--$100~\mathrm{fm}/c$, indicating a similar preferred discrete coalescence time for these nuclei. In all three cases, the minimum $\chi^2/\mathrm{NDF}$ is below unity, corresponding to a satisfactory description of the data.

In contrast, ${}^{4}\mathrm{He}$ exhibits a preference for substantially earlier coalescence times. When all coalescence channels are included, the minimum $\chi^2/\mathrm{NDF}$ occurs at $t_{\mathrm{coal}}\approx 50~\mathrm{fm}/c$, while the direct $p+p+n+n$ channel alone favors $t_{\mathrm{coal}}\approx 40~\mathrm{fm}/c$. The minimum $\chi^2/\mathrm{NDF}$ is 1.5 when all channels are included, but increases to 18.2 for the direct channel alone. The preferred discrete coalescence time for ${}^{4}\mathrm{He}$ is therefore significantly smaller than that inferred from $d$, $t$, and ${}^{3}\mathrm{He}$.

\begin{figure}[htbp]
    \centering
    \includegraphics[width=0.95\columnwidth]{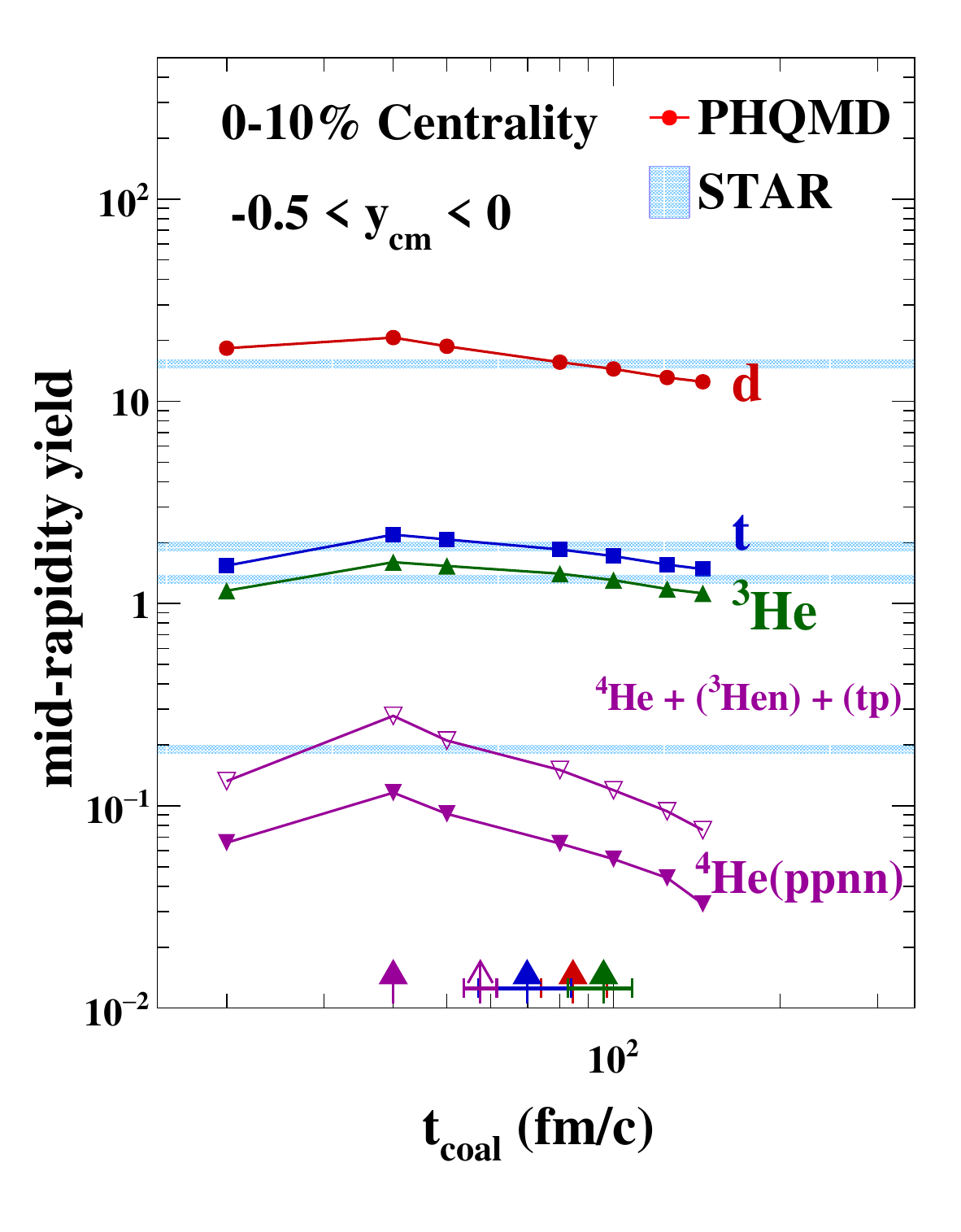}
    \caption{Time dependence of the yields integrated over the rapidity interval $-0.5<y_{\rm cm}<0$ for deuterons, tritons, ${}^{3}\mathrm{He}$, and ${}^{4}\mathrm{He}$ in Au+Au collisions at $\sqrt{s_{NN}}=3$ GeV for the 0--10\% centrality. The colored curves show the PHQMD+coalescence results as a function of the coalescence time $t_{\mathrm{coal}}$, while the light-blue bands indicate the STAR values with their experimental uncertainties. For ${}^{4}\mathrm{He}$, both the direct four-body channel $p+p+n+n$ and the multi-channel result including ${}^{3}\mathrm{He}+n$ and $t+p$ are shown. The arrows at the bottom indicate the matched $t_{\mathrm{coal}}$ values, obtained by interpolating the model curves to the corresponding STAR values. The horizontal bars denote the $1\sigma$ intervals on $t_{\mathrm{coal}}$ obtained from the intersections of the model curves with the upper and lower bounds of the experimental uncertainty.}
    \label{fig:fig6_dndy_time_LN}
\end{figure}

Figure~\ref{fig:fig6_dndy_time_LN} presents the time dependence of the yields integrated over the rapidity interval $-0.5<y_{\rm cm}<0$, and compared with the STAR values. 
The intersections between the model curves and the measured yields provide an estimate of the matched coalescence times. The associated uncertainty intervals are estimated from the intersections with the upper and lower bounds of the experimental uncertainty bands.

For deuterons, tritons, and ${}^{3}\mathrm{He}$, the model curves intersect the STAR uncertainty bands in a similar $t_{\mathrm{coal}}$ range of approximately 70--96~fm/$c$, indicating comparable matched coalescence times for these nuclei.
This behavior is consistent with the preferred discrete $t_{\mathrm{coal}}$ values obtained from the $\chi^2/\mathrm{NDF}$ analysis shown in Figure~\ref{fig:fig5_chi2ndf_LN}. A different trend is observed for ${}^{4}\mathrm{He}$. The direct $p+p+n+n$ channel alone underpredicts the measured yield throughout the studied time range. Including the additional ${}^{3}\mathrm{He}+n$ and $t+p$ channels substantially increases the calculated yield. The resulting multi-channel calculation overlaps with the STAR uncertainty band around $t_{\mathrm{coal}}\approx57 \pm 4$~fm/$c$, indicating a matched coalescence time significantly earlier than that inferred from $d$, $t$, and ${}^{3}\mathrm{He}$.

\begin{figure}[htbp]
    \centering
    \includegraphics[width=1.0\columnwidth]{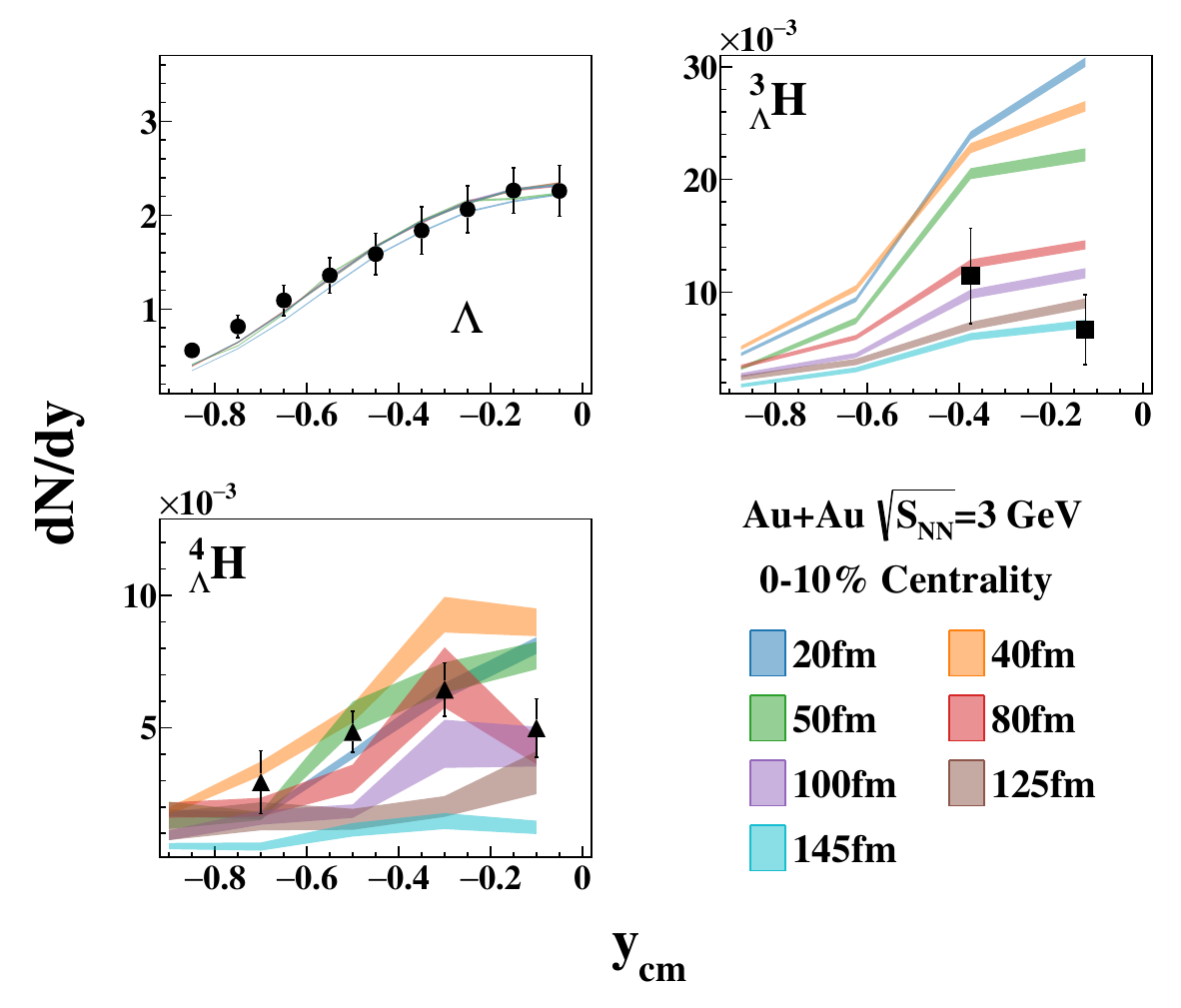}
    \caption{Rapidity distributions, $dN/dy$, of $\Lambda$, ${}^{3}_{\Lambda}\mathrm{H}$, and ${}^{4}_{\Lambda}\mathrm{H}$ in Au+Au collisions at $\sqrt{s_{NN}}=3$ GeV for the 0-10\% centrality. The colored bands show the PHQMD+coalescence results obtained with different $t_{\mathrm{coal}}$, while the symbols denote the STAR data. The hypernuclei data are corrected for the corresponding branching ratios~\cite{Adamczyk2018HypertritonLifetime,Juric1973HypernucleiBinding,Gongleton1992Hypertriton,Kamada1998HypertritonDecay,Outa1998Lambda4Decay}. For ${}^{4}_{\Lambda}\mathrm{H}$, the calculated yield includes both the direct four-body coalescence channel and the additional ${}^{3}_{\Lambda}\mathrm{H}+n$ channel.}
    \label{fig:fig7_dndy_HN}
\end{figure}

Figure~\ref{fig:fig7_dndy_HN} shows the rapidity distribution of $\Lambda$, ${}^{3}_{\Lambda}\mathrm{H}$, and ${}^{4}_{\Lambda}\mathrm{H}$ for the 0-10\% centrality. The PHQMD+coalescence results are compared with the STAR data~\cite{STARHypernucleiLifetime2022,STAR:2024znc} for different values of $t_{\mathrm{coal}}$. 
For $\Lambda$ hyperons, the calculated rapidity distributions exhibit only a weak dependence on $t_{\mathrm{coal}}$ and provide a good description of the STAR measurements.

In contrast, the hypernuclear yields exhibit a much stronger sensitivity to $t_{\mathrm{coal}}$. For both ${}^{3}_{\Lambda}\mathrm{H}$ and ${}^{4}_{\Lambda}\mathrm{H}$, the calculated yields decrease as $t_{\mathrm{coal}}$ increases. 
Similar to ${}^{4}\mathrm{He}$, the ${}^{4}_{\Lambda}\mathrm{H}$ yield receives contributions from both the direct four-body coalescence channel and the ${}^{3}_{\Lambda}\mathrm{H}+n$ channel.

\begin{figure}[htbp]
    \centering
    \includegraphics[width=1\columnwidth]{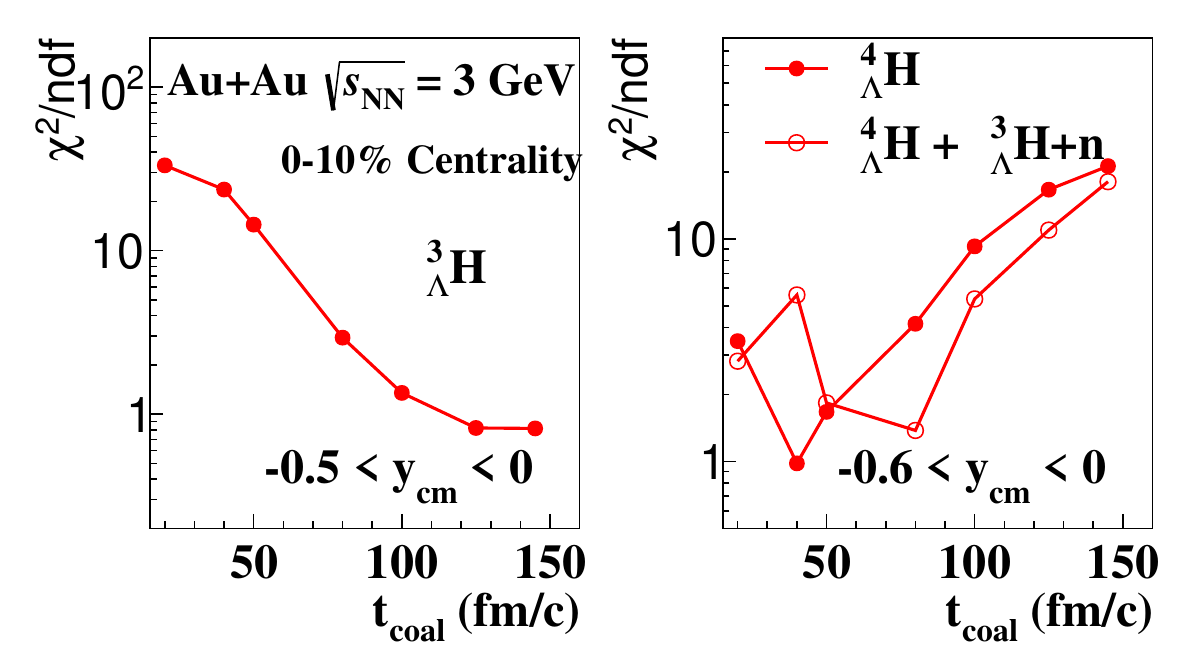}
    \caption{$\chi^2/\mathrm{NDF}$ as a function of the $t_{\mathrm{coal}}$ for hypernucleus production in Au+Au collisions at $\sqrt{s_{NN}}=3$ GeV for the 0-10\% centrality. The left panel shows the result for ${}^{3}_{\Lambda}\mathrm{H}$ in the rapidity interval $-0.5<y_{\mathrm{cm}}<0$. The right panel shows the result for ${}^{4}_{\Lambda}\mathrm{H}$ in the rapidity interval $-0.6<y_{\mathrm{cm}}<0.0$. For ${}^{4}_{\Lambda}\mathrm{H}$, the filled markers correspond to the direct four-body coalescence channel, while the open markers include the additional ${}^{3}_{\Lambda}\mathrm{H}+n$ channel. The $\chi^2/\mathrm{NDF}$ values are calculated by comparing the PHQMD+coalescence rapidity distributions with the STAR data.}
    \label{fig:fig8_chi2ndf_HN}
\end{figure}

The corresponding comparison for hypernuclei is shown in Figure~\ref{fig:fig8_chi2ndf_HN}, where the $\chi^2/\mathrm{NDF}$ values are plotted as a function of $t_{\mathrm{coal}}$ for ${}^{3}_{\Lambda}\mathrm{H}$ and ${}^{4}_{\Lambda}\mathrm{H}$. As in the case of light nuclei, the preferred discrete $t_{\mathrm{coal}}$ is identified from the minimum of the $\chi^2/\mathrm{NDF}$ distribution, where the model best agrees with the experimental data.

For ${}^{3}_{\Lambda}\mathrm{H}$, the minimum $\chi^2/\mathrm{NDF}$ is obtained at $t_{\mathrm{coal}}\approx 125~\mathrm{fm}/c$. In contrast, ${}^{4}_{\Lambda}\mathrm{H}$ favors earlier coalescence times, with minima at approximately $40~\mathrm{fm}/c$ for the direct four-body channel and $80~\mathrm{fm}/c$ when the additional ${}^{3}_{\Lambda}\mathrm{H}+n$ channel is included. Both calculations yield small minimum $\chi^2/\mathrm{NDF}$ values and provide a reasonable description of the data.

\begin{figure}[htbp]
    \centering
    \includegraphics[width=0.95\columnwidth]{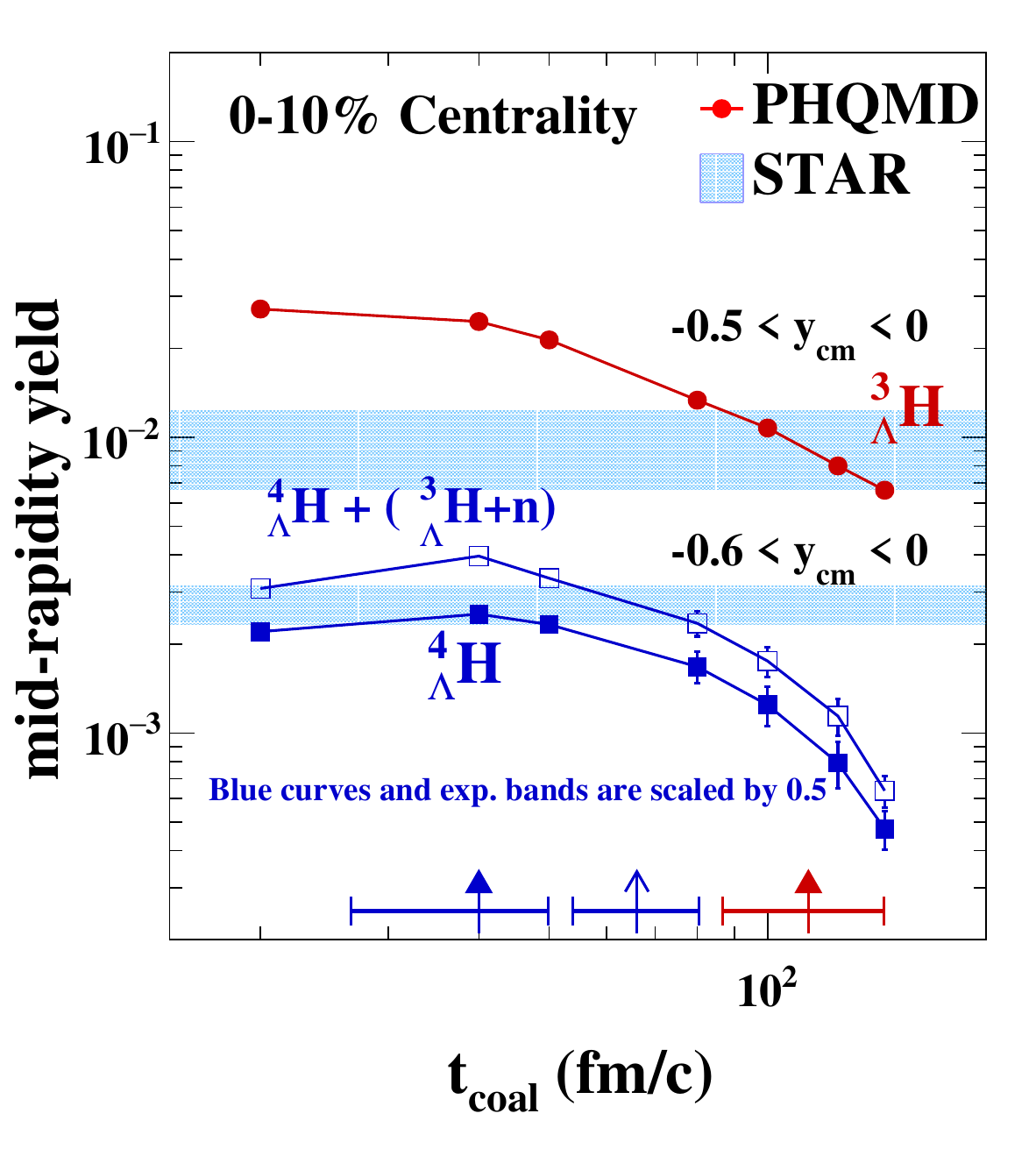}
    \caption{Time dependence of the yields integrated over the mid-rapidity, for ${}^{3}_{\Lambda}\mathrm{H}$, and ${}^{4}_{\Lambda}\mathrm{H}$ in Au+Au collisions at $\sqrt{s_{NN}}=3$ GeV for the 0-10\% centrality and mid-rapidity. The colored curves show the PHQMD+coalescence results as a function of the coalescence time $t_{\mathrm{coal}}$, while the light-blue bands indicate the STAR values with their experimental uncertainties. For ${}^{4}_{\Lambda}\mathrm{H}$, both the direct four-body channel $p+n+n+\Lambda$ and the multi-channel result including the additional ${}^{3}_{\Lambda}\mathrm{H}+n$ channel are shown; the corresponding blue model curves and STAR band are scaled by a factor of 0.5 for visibility. The arrows at the bottom indicate the matched $t_{\mathrm{coal}}$ values, obtained by interpolating the model curves to the corresponding STAR values. The horizontal bars denote the $1\sigma$ intervals on $t_{\mathrm{coal}}$ obtained from the intersections of the model curves with the upper and lower bounds of the experimental uncertainty. For the direct ${}^{4}_{\Lambda}\mathrm{H}$ channel, the bar indicates the overlap with the STAR uncertainty band.}
    \label{fig:fig9_dndy_time_HN}
\end{figure}

Figure~\ref{fig:fig9_dndy_time_HN} presents the integrated hypernuclear yields as a function of $t_{\mathrm{coal}}$. For ${}^{3}_{\Lambda}\mathrm{H}$, the calculated yield decreases with increasing $t_{\mathrm{coal}}$ and reaches the STAR measurement only at 114~fm/$c$, consistent with the preferred discrete $t_{\mathrm{coal}}\approx 125~\mathrm{fm}/c$ obtained from the $\chi^2/\mathrm{NDF}$ analysis.

For ${}^{4}_{\Lambda}\mathrm{H}$, both the direct four-body channel and the multi-channel calculation including the additional ${}^{3}_{\Lambda}\mathrm{H}+n$ contribution intersect the STAR uncertainty band. The direct channel reaches
the STAR band at an earlier time, at $t_{\mathrm{coal}}=40^{+10}_{-13}~\mathrm{fm}/c$, whereas the multi-channel calculation intersects it at a later time, at $66^{+14}_{-13}~\mathrm{fm}/c$. This is consistent with the preferred discrete $t_{\mathrm{coal}}$ obtained from the $\chi^2/\mathrm{NDF}$ analysis in Fig.~\ref{fig:fig8_chi2ndf_HN}.

\begin{figure*}[htbp]
    \centering
    \includegraphics[width=0.95\textwidth]{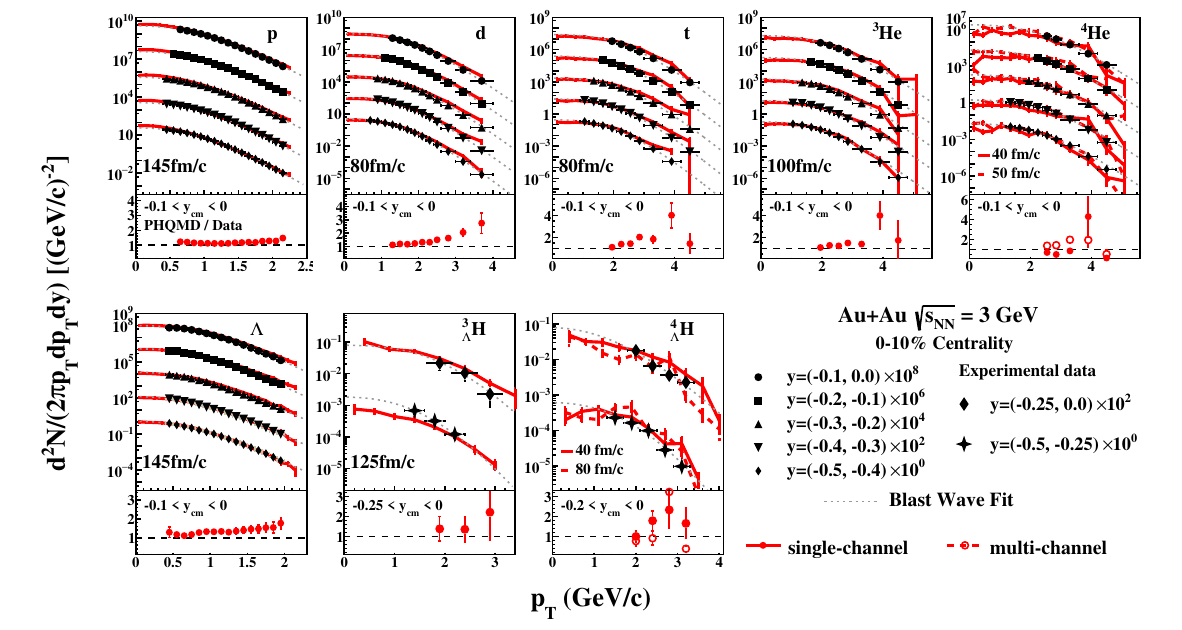}
    \caption{Transverse-momentum spectra of protons, light nuclei, $\Lambda$ hyperons, and hypernuclei in Au+Au collisions at $\sqrt{s_{NN}}=3$~GeV for the 0-10\% centrality. The symbols denote the STAR data in different rapidity intervals, while the red curves show the PHQMD+coalescence results evaluated at the preferred discrete $t_{\mathrm{coal}}$ for each cluster, as determined from the minimum of the $\chi^2/\mathrm{NDF}$ distributions. The lower panels show the ratios of the model results to the data. The points in the ratio panels correspond to the most central mid-rapidity interval. For ${}^{4}\mathrm{He}$ and ${}^{4}_{\Lambda}\mathrm{H}$, both the single-channel and multi-channel results are shown. The gray dashed curves represent the blast-wave fits to the experimental spectra.}
    \label{fig:fig10_pt_LN_HN}
\end{figure*}

Using the preferred discrete $t_{\mathrm{coal}}$ values extracted from the $\chi^2/\mathrm{NDF}$ analysis, Figure~\ref{fig:fig10_pt_LN_HN} compares the transverse-momentum spectra from PHQMD+coalescence with the STAR measurements for protons, light nuclei, $\Lambda$ hyperons, and hypernuclei. The preferred discrete time for each cluster is indicated in the corresponding panel. This comparison provides a differential test of whether the evolution time selected from the rapidity-distribution analysis can also describe the measured $p_{\rm T}$-dependent spectra.

For protons and $\Lambda$ hyperons, the calculated spectra show a reasonable consistency with the data over the measured rapidity intervals. For light nuclei, the spectra obtained at the preferred $t_{\mathrm{coal}}$ values capture the overall magnitude and slope of the STAR data. For ${}^{4}\mathrm{He}$, the single-channel and multi-channel results are shown at their respective preferred $t_{\mathrm{coal}}$ values. Including the additional ${}^{3}\mathrm{He}+n$ and $t+p$ channels changes the calculated spectrum and leads to a closer comparison with the STAR data.

The hypernuclear spectra show a similar overall behavior. The ${}^{3}_{\Lambda}\mathrm{H}$ spectrum at the preferred $t_{\mathrm{coal}}$ provides a reasonable description of the measured points within the current uncertainties. For ${}^{4}_{\Lambda}\mathrm{H}$, the direct and multi-channel results are shown at their respective preferred $t_{\mathrm{coal}}$ values. 
Both calculations give a comparable description of the data, although they favor different $t_{\mathrm{coal}}$ values.

Overall, the preferred discrete $t_{\mathrm{coal}}$ values extracted from the integrated-yield comparisons provide a reasonable description of the measured spectra of light nuclei and hypernuclei in the lower-transverse-momentum region, $p_{\rm
T}/A \lesssim 1~\mathrm{GeV}/c$. At higher $p_{\rm T}/A$, noticeable deviations between the model calculations and the experimental data for $d$ and $t$ remain. A detailed investigation of the high-$p_{\rm T}$ behavior is beyond the scope
of the present study. In the unmeasured low-$p_{\rm T}$ region, the model spectra exhibit a similar overall behavior to the blast-wave extrapolations used in the experimental analysis.

\begin{figure}[htbp]
    \centering
    \includegraphics[width=1\columnwidth]{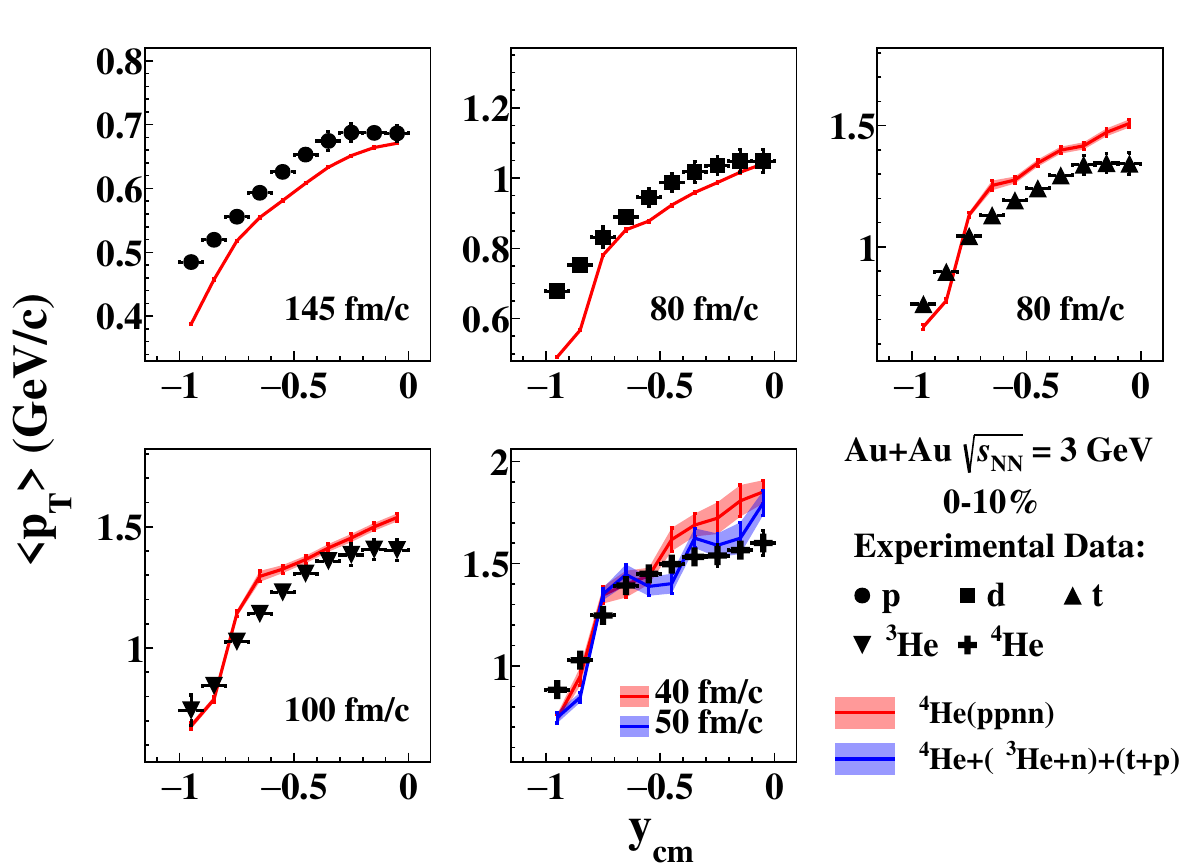}
    \caption{Rapidity dependence of the mean transverse momentum, $\langle p_T\rangle$, for protons, light nuclei, $\Lambda$ hyperons, and hypernuclei in Au+Au collisions at $\sqrt{s_{NN}}=3$ GeV for the 0-10\% centrality. The symbols denote the STAR data, while the curves show the PHQMD+coalescence results evaluated at the preferred discrete $t_{\mathrm{coal}}$ values for each particle species, as determined from the minima of the $\chi^2/\mathrm{NDF}$ distributions. For ${}^{4}\mathrm{He}$, both the single-channel result from the direct $p+p+n+n$ channel and the multi-channel result including the additional ${}^{3}\mathrm{He}+n$ and $t+p$ channels are shown.}
    \label{fig:fig11_meanpt_LN}
\end{figure}

Figure~\ref{fig:fig11_meanpt_LN} compares the calculated $\langle p_T \rangle$ values at the preferred discrete $t_{\mathrm{coal}}$ with the STAR measurements. For protons and light nuclei, the PHQMD+coalescence calculation reproduces the overall rapidity dependence of $\langle p_T \rangle$, including the increase from backward rapidity toward mid-rapidity. While quantitative differences remain in some rapidity intervals, particularly for the heavier clusters, no large discrepancy is observed between the calculations and the data.

It should be noted that the experimental $\langle p_T \rangle$ values depend on the extrapolation into the unmeasured low-$p_T$ region. Therefore, the overall agreement between the model calculations and the measurements provides additional support that the extrapolated contribution is of a reasonable magnitude.

\begin{figure}[htbp]
    \centering
    \includegraphics[width=0.95\columnwidth]{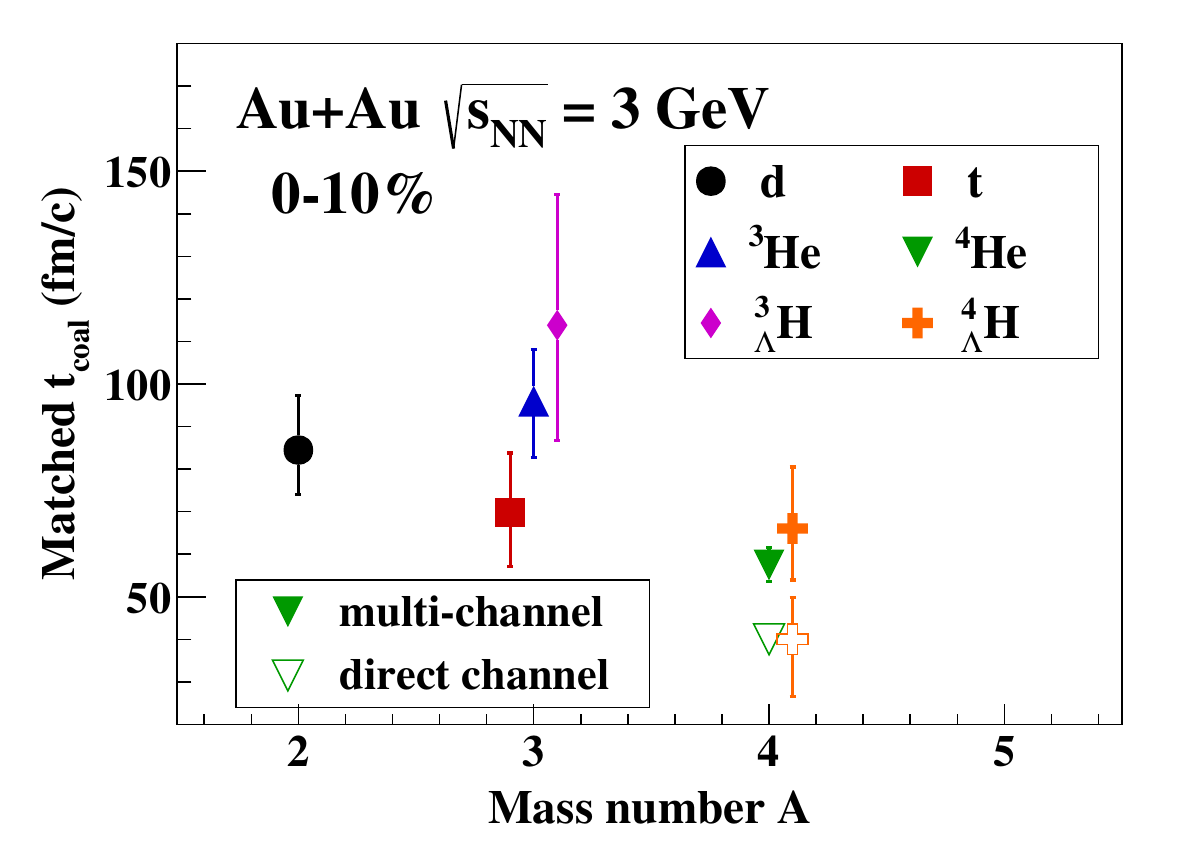}
    \caption{Matched coalescence time \(t_{\mathrm{coal}}\) as a function of mass number \(A\) for light nuclei and hypernuclei in Au+Au collisions at \(\sqrt{s_{NN}}=3\) GeV for the 0-10\% centrality. The central values are obtained from the interpolated intersections of the PHQMD+coalescence integrated-yield curves with the STAR values. The vertical error bars indicate the corresponding time intervals estimated from the intersections with the upper and lower edges of the STAR uncertainty bands.  For ${}^{4}\mathrm{He}$ and ${}^{4}_{\Lambda}\mathrm{H}$, open and filled markers denote the direct-channel and multi-channel results, respectively.}
    \label{fig:fig12_Time_A}
\end{figure}

The matched coalescence times extracted from the integrated-yield comparisons are summarized in Fig.~\ref{fig:fig12_Time_A}. The extracted values indicate a species-dependent pattern. The largest matched coalescence times are obtained for ${}^{3}_{\Lambda}\mathrm{H}$, followed by the non-strange $A=2$ and $A=3$ nuclei. The ${}^{4}\mathrm{He}$ and ${}^{4}_{\Lambda}\mathrm{H}$ are associated with comparatively smaller matched coalescence times. For the $A=4$ systems, the direct-channel and multi-channel results are shown separately, illustrating the sensitivity of the matched coalescence time to the assumed formation channels.

\begin{figure}[htbp]
    \centering
    \includegraphics[width=0.95\columnwidth]{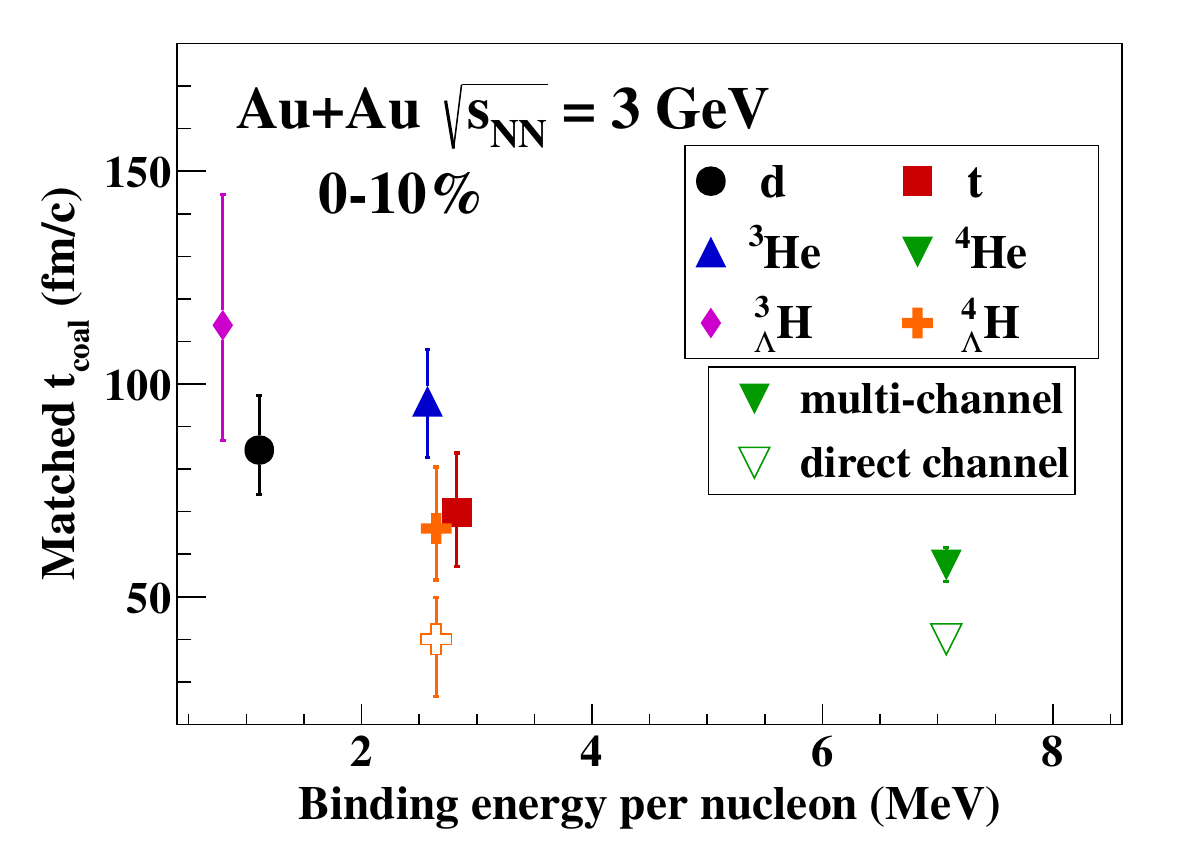}
    \caption{Matched coalescence time $t_{\mathrm{coal}}$ as a function of binding energy per nucleon for light nuclei and hypernuclei in Au+Au collisions at $\sqrt{s_{NN}}=3~\mathrm{GeV}$ for the 0--10\% centrality. The central values are obtained from the interpolated intersections of the PHQMD+coalescence integrated-yield curves with the STAR values. The vertical error bars indicate the corresponding time intervals estimated from the intersections with the upper and lower edges of the STAR uncertainty bands. For ${}^{4}\mathrm{He}$ and ${}^{4}_{\Lambda}\mathrm{H}$, open and filled markers denote the direct-channel and multi-channel results, respectively.}
    \label{fig:fig15_Time_BE}
\end{figure}

Figure~\ref{fig:fig15_Time_BE} presents the extracted matched coalescence times as a function of the binding energy per nucleon. Compared with the mass-number representation in Fig.~\ref{fig:fig12_Time_A}, this presentation allows the results to be examined in terms of nuclear binding strength. Using the multi-channel results for the $A=4$ systems, the extracted matched coalescence times are compatible, within uncertainties, with a decreasing trend as the binding energy per nucleon increases. Such a trend is less evident when only the direct coalescence channels are considered. However, for ${}^{4}\mathrm{He}$, the direct-channel calculation is unable to reproduce the measured yield, indicating that additional formation channels may play an important role. Future higher-statistics measurements, particularly of heavier nuclei and hypernuclei such as ${}^{5}_{\Lambda}\mathrm{He}$, will provide important constraints on the role of multi-channel formation mechanisms and the connection between cluster binding and the coalescence time.

\subsection{Predictions for ${}^{5}_{\Lambda}\mathrm{He}$ and ${}^{5}_{\Lambda\Lambda}\mathrm{He}$ production}
\begin{figure}[htbp]
    \centering
    \includegraphics[width=0.95\columnwidth]{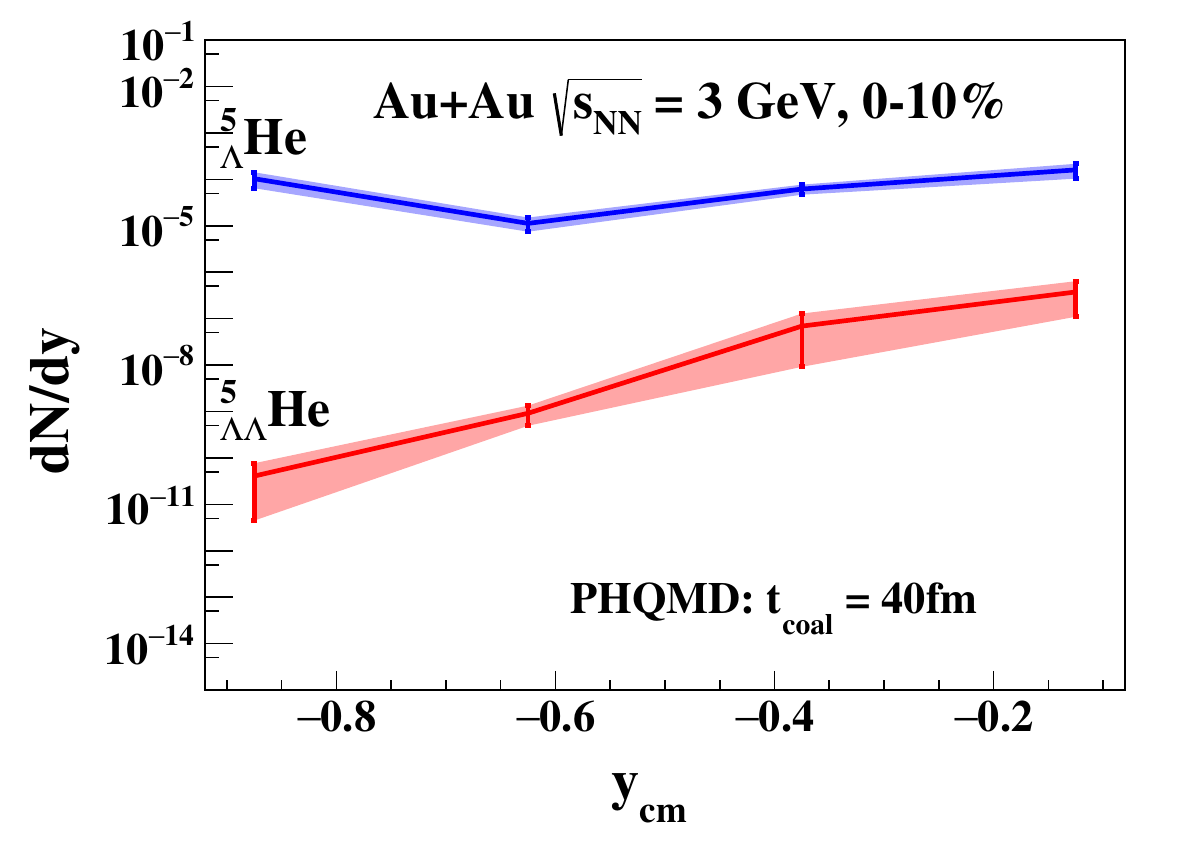}
    \caption{Rapidity distributions, $dN/dy$, of ${}^{5}_{\Lambda}\mathrm{He}$ and ${}^{5}_{\Lambda\Lambda}\mathrm{He}$ predicted by PHQMD+coalescence in Au+Au collisions at $\sqrt{s_{NN}}=3$ GeV for the 0-10\% centrality. The results are shown for $t_{\mathrm{coal}}=40~\mathrm{fm}/c$.}
    \label{fig:fig13_He5L_He5LL}
\end{figure}

Beyond ${}^{3}_{\Lambda}\mathrm{H}$ and ${}^{4}_{\Lambda}\mathrm{H}$, the study of ${}^{5}_{\Lambda}\mathrm{He}$ and double strangeness cluster ${}^{5}_{\Lambda\Lambda}\mathrm{He}$ provides access to more deeply bound hypernuclear systems and offers additional constraints on the $\Lambda N$ and $\Lambda\Lambda$ interactions. Their production yields and kinematic properties therefore constitute sensitive probes of both hypernuclear structure and the formation mechanisms of strange nuclei in heavy-ion collisions.
Therefore, we provide predictions for the rapidity distributions of heavier hypernuclei, ${}^{5}_{\Lambda}\mathrm{He}$ and ${}^{5}_{\Lambda\Lambda}\mathrm{He}$, as shown in Figure~\ref{fig:fig13_He5L_He5LL}. 
Following the trend observed in Figure~\ref{fig:fig12_Time_A}, we choose a early coalescence time, $t_{\mathrm{coal}}=40~\mathrm{fm}/c$, for the 0-10\% centrality.
The predicted ${}^{5}_{\Lambda}\mathrm{He}$ yield is several orders of magnitude larger than that of ${}^{5}_{\Lambda\Lambda}\mathrm{He}$, reflecting the additional suppression associated with double-$\Lambda$ hypernucleus formation. 
These results provide reference estimates for future experimental searches for heavier double-$\Lambda$ hypernuclei at $\sqrt{s_{NN}}=3$ GeV.

\section{Summary}

In this work, we investigated the formation of light nuclei and hypernuclei in Au+Au collisions at $\sqrt{s_{NN}}=3$ GeV within a PHQMD+coalescence framework. The coalescence probabilities were evaluated using Wigner phase-space densities constructed from realistic few-body wave functions, and the coalescence time was implemented through the hadronic evolution cutoff time $t_{\mathrm{coal}}$.

The calculated cluster yields show a clear dependence on $t_{\mathrm{coal}}$. Comparisons with STAR nuclei and hypernuclei yield measurements were used to extract matched coalescence times for different cluster species. The extracted times exhibit a species-dependent pattern: the $A=2$ and $A=3$ clusters are generally matched at later times, whereas the $A=4$ systems favor earlier times. 
This imply that they are formed in a denser environment where correlations are stronger. In a dilute late-stage system, it is difficult for four nucleons to simultaneously satisfy the coalescence condition. Consequently, heavier clusters may retain information from an earlier\cite{Sun:2015Li,Wang:2026Clustered}, high-density phase of the reaction and thus serve as sensitive probes of the high-baryon-density region. 
We further find that cluster--nucleon formation channels may play an important role in reproducing the measured yields of ${}^{4}\mathrm{He}$. Using the corresponding multi-channel results for ${}^{4}\mathrm{He}$ and ${}^{4}_{\Lambda}\mathrm{H}$, the extracted matched coalescence times are compatible, within uncertainties, with a tendency toward earlier formation for more strongly bound nuclei and hypernuclei.

\FloatBarrier
\begin{acknowledgments}
This work is supported in part by the National Natural Science Foundation of China under Grant No. 12375134, the National Key Research and Development Program of China (Grant No. 2024YFE0110103 and 2024YFA1611003). JX is supported by Helmholtz Research Academy Hesse for FAIR (HFHF).
\end{acknowledgments}


\bibliography{main}

\appendix

\end{document}